# Is the standard quantum mechanics a completely nondeterministic theory?

Habibollah Razmi [(1)] and Jafar Bahreini [(2)]

(1) Department of Physics, University of Qom, Qom, I. R. Iran.

(2) Department of Philosophy, University of Qom, Qom, I. R. Iran.

(1) razmi@qom.ac.ir (2) jf.bah1400@gmail.com

## Abstract

It is argued that although quantum theory isn't an absolutely deterministic theory, it is partially deterministic. The approach followed here is in the framework of the standard (Copenhagen interpretation of) quantum mechanics without any additional assumption or alternative interpretation. The argument is based on the conceptual meaning of determinism and by means of some well-known phenomena in the quantum world (measurement of a quantum particle spin, energy values and spatial states of an electron in atoms, the so-called wavefunction collapse problem) which are usually considered in rejecting determinism.

**Keywords**: Determinism; indeterminism; Quantum Mechanics

## 1. Introduction

Concept of determinism has been under consideration from the ancient history of science to the classical physics era and then modern physics stage. It has provided special attentions of scientists and philosophers by formation and development of modern quantum mechanics. In this letter, to be considered mostly by physicists than philosophers, without referring to a huge number of books and articles on the subject of determinism which are easily in access by a simple search via internet (e.g. from Laplace famous definition and consideration of determinism in classical physics [1] to modern researches by Earman [2-3] and references therein, Butterfield [4], Müller and Placek [5]), we want to argue that the standard quantum mechanics (the so-called Copenhagen interpretation of quantum theory) isn't a completely nondeterministic and merely chancy theory. Our approach is based on considering and discussing about some well-known phenomena in the quantum world which are appeared to be completely nondeterministic but it can be argued that they aren't absolutely indeterminate and chancy but partially and incompletely deterministic.

## 2. Determinism in classical mechanics

Although some ones, using very special rarely found examples (e.g. from chaotic systems), have tried to show that classical physics deals with nondeterministic phenomena [2-4], it is well-known that in classical physics the state of any particle/body/system at any instant of time is a uniquely determined effect of its past and the cause of its future with a one to one correspondence in a completely/absolutely causal deterministic manner; this can be seen both mathematically and intuitively via the second law of motion of Newtonian dynamics in which a second order differential equation governs the motion of a body with a unique solution such that when for example its position and velocity known at an instance of time, its status is completely known/determined in the past/future. Existence of some perturbations or tolerances in knowing/determining the past/future isn't due to the nondeterministic character of the theory but due to lack of our information about the initial conditions and influences of other bodies/systems or some generalizations of the theory due to for example perturbative and relativistic effects on the motion of the earth in its orbit around the sun for which we can extract a thousands of years past/present/future exact calendar.

**2. Determinism in quantum mechanics**

In quantum mechanics, the state of any quantum object is under government of the Schrödinger equation [6] and the corresponding 'causal' time evolution and thus some ones claim that quantum mechanics is a deterministic theory [2-4]. Of course, this can at most show that the theory is causal but not absolutely deterministic because the position and other physical properties of the object are specified by the square value of the wavefunction as a probability value (Born's postulate [7]). The Heisenberg uncertainty principle [8] and the probabilistic character of the standard quantum theory are the fundamental bases for a number of (sometimes an infinite number of) possible states for a quantum object with indeterminate values for its physical quantities as its position, momentum, energy, etc.

**3. Can determinism be denied completely by quantum mechanics?**

As an example from quantum mechanics, consider an electron (not a statistical beam but just one electron) in an un-polarized state for which an experiment (such as Stern-Gerlach apparatus [9-11]) has been designed to measure its spin. The standard quantum mechanics teaches us that if one does the experiment, he shall finds out an up (+1/2) outcome with a probability of %50 and a down (-1/2) outcome with a probability of %50. This shows there is a probable and inderemistic situation; moreover, repetition of the experiment with the same conditions doesn't result in the same result as before. Well, even if we accept this example without considering any statistical substratum and without any doubt about the guarantee of repetition of the experiment as the same as before, there are still a kind of "predictability" (determinism) in this experiment and not an absolute indeterminism; because, clearly confirmed by the standard quantum mechanics, we can "predict" the resulting outcome is up or down and not another unknown result (e.g. +1/3 or -1/7 or other infinite possible results). In other words, the result of the experiment is pre-determined in a specific area but not as an exact point.

Moreover, as we know, the time evolution of the spin state of the electron is under government of the Schrödinger differential equation in which any present state is a function of the past state and the initial state for the future. So, although the experiment isn't classically/mechanically/absolutely deterministic, it is a causal deterministic experiment in an incomplete manner. As another example consider the case of dealing with infinite energy states of a particle in a box. Although there are an infinite possible discrete energy values depending on some constants corresponding to the geometry of the box, all them are under the government of a specific mathematical formula and not an arbitrary chancy value; thus, the results are pre-determined in a specific "area" or "domain". This is true for the case of the energy levels of the electrons in atoms which are well-known in the literature of quantum mechanics as real examples of the quantum micro world. Here again although there are an infinite possible energy levels, all of them are constrained based on fixed mathematical formulae; this means although there isn't a classical complete deterministic situation, there isn't a complete nondeterministic situation either (i.e. an incomplete determinism governs the phenomenon). One of the most successful application of the standard quantum mechanics is in explaining the energy levels and the status of the electronic structure of the Hydrogen atom; as we know, although there are infinite possible/probable radii and energy levels for the electron in the Hydrogen atom, a special mathematically pre-determined ordered of radii with regular values of energies with an orderly principle quantum number are allowed and not any arbitrary mathematical function with a merely chancy value.

In the very important subject of measurement and the wavefunction collapse problem [12-14] which is usually considered as one of the most important examples reasoning on the denial of determinism in quantum mechanics, it is clear that the final (even sometimes with an infinite number of) results and states of collapse are specified (determined) in a pre-known region of places/states not in a merely chancy and infinitely unknown status. For example, corresponding to the above-mentioned example of spin 1/2 particles, the spin state collapse occurs within the domain of the two results +1/2 or -1/2 and not any other chancy results as +1/3 or -1/3 etc.; for the example of the measurement of the energy level state of an atomic electron (e.g. hydrogen atom), although there are an infinite number of the final states of the collapse due to the measurement, all these final states restricted to determined radii with specific energy values and not any infinitely unknown chancy mathematical function of space with an arbitrary random value of energy. Yes, there isn't a completely forced and constrained final state with a one to one correspondence between the initial (before measurement) and final (after collapse) states as what happens in the classical physics; but there is still a local correspondence between the initial and final states with a predetermined range/region of final states. There is both not a perfectly absolute final state not a perfectly chancy and undetermined final state.

## 6. Conclusion and discussion

We conclude with this fact that determinism cannot be completely denied in quantum theory. The standard (Copenhagen interpretation of) quantum mechanics is a deterministic theory but in an incomplete manner. It is noticeable to mention that there is a fundamental difference between what we have considered in this letter and some other already well-known alternative interpretation of quantum mechanics as those based on hidden variable theories (the most famous one is Bohmian Quantum mechanics [15-16]) or the so-called many worlds interpretation of quantum theory [17]. Indeed, such alternative theories and approaches have been presented to save determinism/causality and realism at an underside (substratum) level for the standard formulation of quantum theory while we have tried to explain about this fact that even Copenhagen interpretation of quantum mechanics doesn't reject determinism completely/absolutely. What is rejected by the standard quantum mechanics is a one to one constrained/forced correspondence (an absolute/complete determinism); but of course, the standard quantum mechanics is a partially deterministic theory. Quantum particles aren't in a completely deterministic state and not in a completely random and free situation but in an intermediate state. Not only at the macroscopic scales but also at the microscopic levels, our world isn't in a completely indeterminate, wanderful, and aimless state.

Our argument here can be simply applied to the advanced level of quantum mechanics as the quantum theory of radiation for some apparently nondeterministic phenomena as radioactive decay or "spontaneous" emission of atoms. As we know, both the decaying of unstable atoms/nuclei and "spontaneous" emission of atoms are due to the quantum fluctuations of the background vacuum with specified/determined channels of decay and final stable states through mathematically restricted transition rules; these are all reason on this fact that although these processes aren't completely/absolutely deterministic with one to one correspondence between the initial and final states, they all partially deterministic in an already known domain and not in an arbitrary complete chancy and indeterminate manner.

Our discussion may correspond to the subject of the particles' consciousness and their "ability" (the important subject of "free will") in choosing a state among other possible states [18]; this needs another commentary research work.